\documentclass[aps,notitlepage,twocolumn,nofootinbib]{revtex4-1}

\usepackage{amsmath}
\usepackage{amsfonts}
\usepackage{amssymb}
\usepackage{physics}
\usepackage{slashed}
\usepackage{simplewick}

\usepackage{siunitx}

\usepackage{graphicx}
\usepackage{float}

\usepackage[caption=false]{subfig}

\usepackage{listings}

\usepackage{tikz}
\usetikzlibrary{calc}
\usetikzlibrary{patterns}
\usetikzlibrary{positioning,decorations.pathmorphing,decorations.markings}
\usetikzlibrary{shapes,arrows}

\usepackage{feynmp-auto}
\usepackage{multirow}

\usepackage{hyperref}

\newcommand{\myAffiliation}{Institute of Physics, National Chiao-Tung University, 1001 Ta-Hsueh Road, Hsinchu 30010, Taiwan}

\usepackage{xcolor}

\begin{document}

\begin{abstract}
We explore the feasibility of determining Mellin moments of the pion's light cone distribution amplitude using the heavy quark operator product expansion (HOPE) method. As the first step of a proof of principle study we pursue a determination of the second Mellin moment. We discuss our choice of kinematics which allows us to successfully extract the moment at low pion momentum. We describe the numerical simulation, and describe the data analysis, which leads us to a preliminary determination of the second Mellin moment in the continuum limit in the quenched approximation as $\expval{\xi^2}=0.19(7)$ in the $\overline{\text{MS}}$ scheme at 2 GeV.
\end{abstract}

\title{A Preliminary Determination of the Second Mellin Moment of the Pion's Distribution Amplitude Using the Heavy Quark Operator Product Expansion}

\author{William Detmold}
\affiliation{Center for Theoretical Physics, Massachusetts Institute of Technology, 77 Massachusetts Avenue, Cambridge, MA 02139, USA}

\author{Anthony V. Grebe}
\email{agrebe@mit.edu, Speaker at APLAT 2020}
\affiliation{Center for Theoretical Physics, Massachusetts Institute of Technology, 77 Massachusetts Avenue, Cambridge, MA 02139, USA}

\author{Issaku Kanamori}
\affiliation{RIKEN Center for Computational Science, Kobe 650-0047, Japan}

\author{C.-J. David Lin}
\affiliation{\myAffiliation}
\affiliation{Centre for High Energy Physics, Chung-Yuan Christian University, Chung-Li, 32032, Taiwan}

\author{Santanu Mondal}
\affiliation{Los Alamos National Laboratory, Theoretical Division T-2, Los Alamos, NM 87545, USA}

\author{Robert J. Perry}
\email{perryrobertjames@gmail.com, Speaker at APLAT 2020}
\affiliation{\myAffiliation}

\author{Yong Zhao}
\affiliation{Physics Department, Brookhaven National Laboratory, Bldg. 510A, Upton, NY 11973, USA}

\collaboration{HOPE Collaboration}

\maketitle

\section{Introduction}

At high energies, exclusive processes in quantum chromodynamics (QCD) may be described with the aid of the so-called light cone distribution amplitudes (LCDAs), convolved with a short distance perturbative kernel~\cite{Chernyak:1977fk,Lepage:1980fj}. These distribution amplitudes contain the non-perturbative information about the process, and are the result of a Fock space truncation where only the lowest, valence Fock state is retained. In this sense, one may consider the pion's light cone distribution amplitude $\phi(x,\mu^2)$ as giving the probability amplitude for converting a pion into a pair of collinear quark and antiquark with longitudinal momentum fraction $x$ and $1-x$, respectively. It is defined via the matrix element
\begin{equation}
\begin{split}
\bra{0}&\overline{\psi}(z_2 n)\slashed{n} \gamma_5 W[z_2 n,z_1 n] \psi(z_1 n)\ket{\pi^+(\mathbf{p})}
\\
&=i f_\pi (p\cdot n)\int_{0}^1 dx \, e^{-i(z_1 x+z_2(1-x))p\cdot n}\phi_\pi(x,\mu^2),
\end{split}
\end{equation}
where $p^\mu$ is the momentum of the pion, $n$ is a light-like vector ($n^2=0$), $z_1$ and $z_2$ are real numbers, $f_\pi=0.132$~\si{\GeV} is the pion decay constant, $\mu^2$ is the renormalization scale and $W[a,b]$ is a Wilson line required to ensure gauge invariance of the matrix element.
%
We take $x$ to be the longitudinal momentum fraction of the $u$-quark in the Fock state $\ket{u\overline{d}}$. Momentum conservation then implies that the $\overline{d}$ quark has momentum fraction $1-x$. In the isospin limit, where the masses of the up and down quarks are degenerate, the light cone distribution amplitude is symmetric under the interchange $x\to 1-x$, that is
\begin{equation}
\phi_\pi(x,\mu^2)=\phi_\pi(1-x,\mu^2).
\end{equation}
We shall assume isospin symmetry in this work.


The value of precise determinations of the LCDA lies in the object's  \textit{process independence}. This allows one to describe many exclusive processes in QCD with the same distribution amplitude convolved with a process dependent perturbative kernel.

Currently, the only \textit{ab-initio} approach to the determination of this object is a numerical computation using lattice QCD. Unfortunately, direct calculation of such light-cone objects is impossible on a Euclidean lattice since the light cone, defined by $z^2=0$, is a reduced to a single point ($z_\text{E}^\mu=0$). Despite this difficulty information about the LCDA may be determined from the lattice in a number of indirect ways. The traditional approach involves a determination of the \textit{Mellin moments} of the LCDA~\cite{Kronfeld:1984zv,Martinelli:1987si}. These are defined by
\begin{equation}
\expval{\xi^n}_{\mu^2}=\int_{-1}^1 d\xi\,  \xi^n\phi(\xi,\mu^2),
\end{equation}
where $\xi=2x-1$ and $x$ is the momentum fraction of the collinear quark anti-quark pair. Noting again the isospin symmetry $x\leftrightarrow(1-x)$ we see that only the even moments may be non-zero for the pion. These moments may be related to local matrix elements which are immediately amenable to a lattice calculation. 
It is possible to write the full distribution amplitude with the knowledge of the Mellin moments alone:
\begin{equation}
\phi(\xi,\mu^2) = \frac{1}{2\pi} \int_{-\infty}^\infty ds \left( \sum_{n=0}^\infty \frac{(is)^n}{n!} \expval{ \xi^n}_{\mu^2}  \right) e^{-is\xi}.
\end{equation}

Unfortunately, the breaking of the full rotation group on the lattice leads to operator mixing and thus power divergences appear in twist-2 operators with spin higher than four~\cite{Gockeler:1996mu}. These power divergences make the determination of the higher moments more difficult. Nevertheless, this approach has been well studied and has yielded results for the first non-trivial moment of the pion and kaon~\cite{Braun:2015axa,Bali:2019dqc}. A number of other proposals in the literature seek to overcome this difficulty~\cite{Braun:2007wv,Bali:2017gfr,Ma:2017pxb,Sufian:2020vzb,Davoudi:2012ya,Ji:2013dva,Ji:2014gla,Ji:2020ect,Alexandrou:2018pbm,Lin:2018pvv,Zhang:2017bzy,Chen:2017gck,Zhang:2020gaj,Radyushkin:2017cyf,Orginos:2017kos,Joo:2020spy,Liu:1993cv,Liang:2019frk,Oehm:2018jvm,Can:2020sxc,Chambers:2017dov}.

While much good work has been done in the extraction of the pion LCDA, it is clear that more must still be done to acquire precise predictions of this object. With this view, it is clearly of interest to explore other proposals for the calculation of the distribution amplitude. One such approach, which we pursue in this work is the so-called heavy quark operator product expansion (HOPE)~\cite{Detmold:2005gg,Detmold:2018kwu}. The HOPE method builds on the conventional operator product expansion (OPE) approach, by performing the numerical simulation with a fictitious heavy quark species, which leads to a number of advantages over the standard treatment~\cite{Detmold:2005gg}. This method allows the extraction of the Mellin moments of the LCDA, and thus in principle allows the reconstruction of the amplitude within a wide range of $x$. In this paper, we discuss the application of the HOPE method to the pion's LCDA. In particular, we discuss kinematic choices which lead to an efficient extraction of the second Mellin moment, and discuss the resulting preliminary extraction of the second Mellin moment.


\section{Summary of the Calculation}


The HOPE method is a multi-step procedure. Thus, before beginning our discussion of the kinematics used and the numerical study we performed, we provide an overview of the calculation.
The starting point of this work is a study of the anti-symmetric version of the matrix element in Minkowski space,
\begin{equation}
T^{\mu\nu}(p,q)=\int d^4z\, e^{iq\cdot z}\bra{0}\mathcal{T}[J_A^\mu(z/2) J_A^\nu(-z/2)]\ket{\pi(\mathbf{p})},
\end{equation}
given by 
\begin{equation}
\begin{split}
U^{\mu\nu}(p,q)&=\frac{1}{2}\bigg(T^{\mu\nu}(p,q)-T^{\nu\mu}(p,q)\bigg)
\\
&=\int d^4z\, e^{iq\cdot z}\bra{0}\mathcal{T}[J_A^{[\mu}(z/2) J_A^{\nu]}(-z/2)]\ket{\pi(\mathbf{p})},
\end{split}
\end{equation}
where the axial-vector current is replaced by the heavy-light flavour changing current:
\begin{equation}
J_A^\mu = \bar \Psi \gamma^\mu \gamma^5 \psi + \bar \psi \gamma^\mu \gamma^5 \Psi,
\label{eq:heavy-light}
\end{equation}
with $\psi$ being the light quark field, and $\Psi$ being the heavy quark field. We note that it is also possible to study the LCDA Mellin moments using the corresponding heavy-light vector current. By applying the OPE to the above matrix element, we can show~\cite{Detmold:2018kwu} that to leading twist, the antisymmetric tensor $U^{\mu\nu}(p,q)$ may be written in the isospin limit as\footnote{Note that Ref.~\cite{Detmold:2018kwu}, uses a normalization for the Mellin moments which differs by a factor of $2^{n}$ from our convention. Ours agrees with the `standard' normalization which allows us to directly compare our result with other determinations of the second Mellin moment.}
\begin{equation}
\begin{split}
U^{\mu\nu}(p,q)&=\frac{2if_\pi  \epsilon^{\mu\nu\alpha\beta}q_\alpha p_\beta}{\tilde{Q}^2}\sum_{n\text{ even}}^\infty\frac{\mathcal{C}_n^{2}(\eta)}{2^n(n+1)}C_W^{(n)}(\tilde{Q}^2)
\\
&\times \expval{\xi^n}\zeta^n+\mathcal{O}(1/\tilde{Q}^3),\label{eq:hope}
\end{split}
\end{equation}
where $f_\pi\approx0.132$~\si{GeV} is the pion decay constant, $\tilde{Q}$, $\eta$ and $\zeta$ are kinematic variables given by
\begin{align}
\tilde{Q}^2&=-q^2-m_\psi^2,
\\
\eta&=\frac{p\cdot q}{\sqrt{p^2 q^2}},
\\
\zeta&=\frac{\sqrt{p^2 q^2}}{\tilde{Q}^2}.
\label{eq:kinematcs}
\end{align}
$C_W^{(n)}$ are the Wilson coefficients, and $\mathcal{C}_n^{2}(\eta)$ are the Gegenbauer polynomials, which arise as a result of resumming target mass effects~\cite{Georgi:1976vf,Nachtmann:1973mr}. 

In order to accurately extract the Mellin moments, one needs to determine the Wilson coefficients beyond zeroth order. Since these Wilson coefficients only account for the ultraviolet effects of QCD, they may be calculated using perturbation theory. The Wilson coefficients may be written
\begin{equation}
C_W^{(n)}(\tilde{Q}^2)=1+\alpha_s c_n^{(1)}+\dots.
\end{equation}
By calculating the matrix element $T^{\mu\nu}(p,q)$ on the lattice, one may then perform a fit to the form of the heavy quark OPE, and thus obtain the Mellin moments of the distribution amplitude. We note that when computing the hadronic tensor, kinematics should be chosen such that we remain in the unphysical region. This requires choosing
\begin{equation}
(p+q)^2<m_{hl}^2\approx (m_\Psi+\Lambda_\text{QCD}),
\end{equation}
where $m_{hl}$ is the mass of the lightest heavy-light meson. This ensures that the analytic continuuation to Minkowski space is may be straightforwardly obtained by the replacement $q_4\to iq_0$. Thus this method is constrained to work in the window
\begin{equation}
\Lambda_\text{QCD}\ll\sqrt{q^2}< m_\Psi\ll \frac{1}{a}
\end{equation}
By performing the calculation at a number of lattice spacings, one may then extrapolate to the continuum in the usual way~\cite{inPreparationTheory}.

\subsection{Efficient Kinematics for an Extraction of the Second Mellin Moment}

\begin{figure}
\centering
\includegraphics[width=\columnwidth]{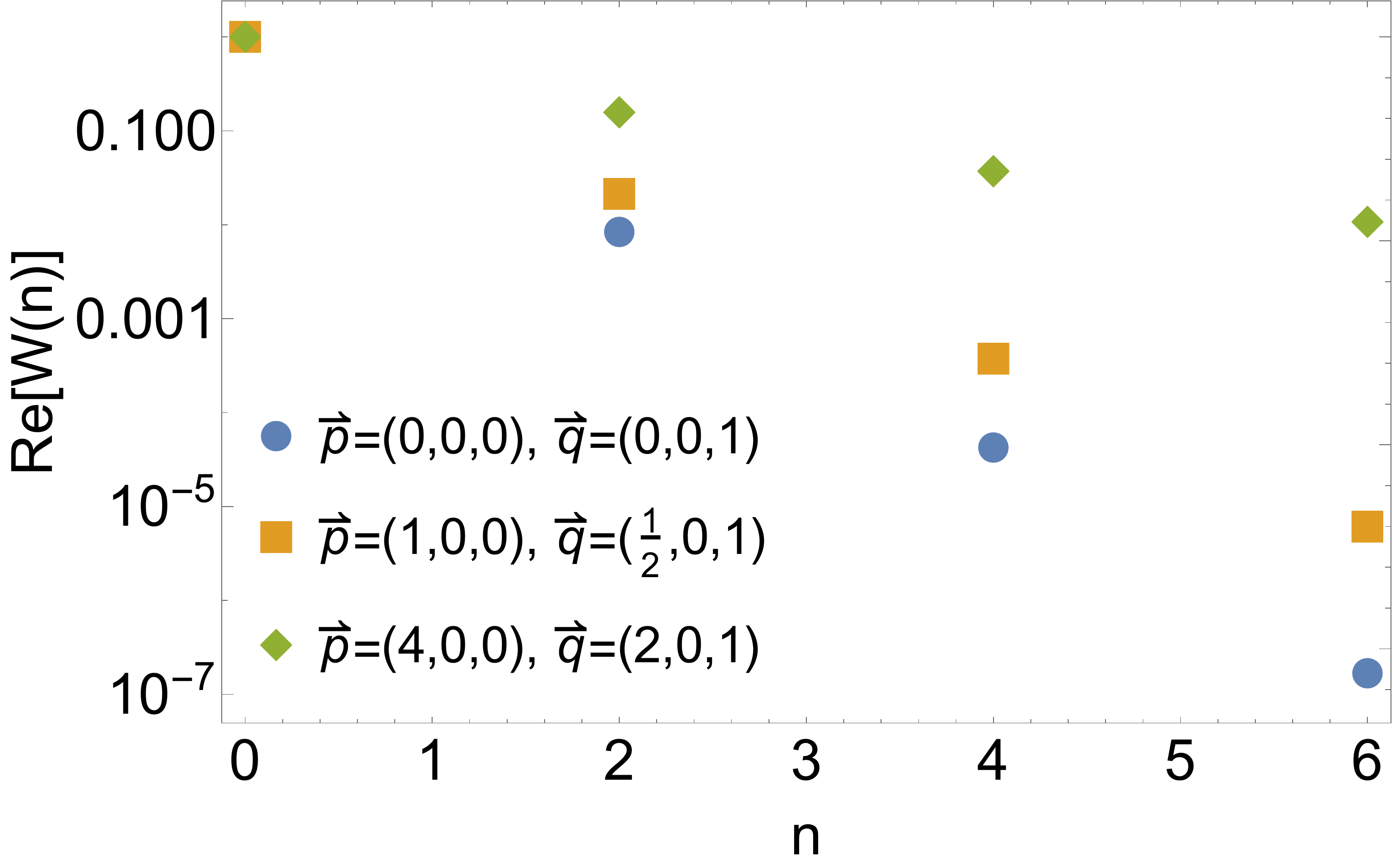}
\includegraphics[width=\columnwidth]{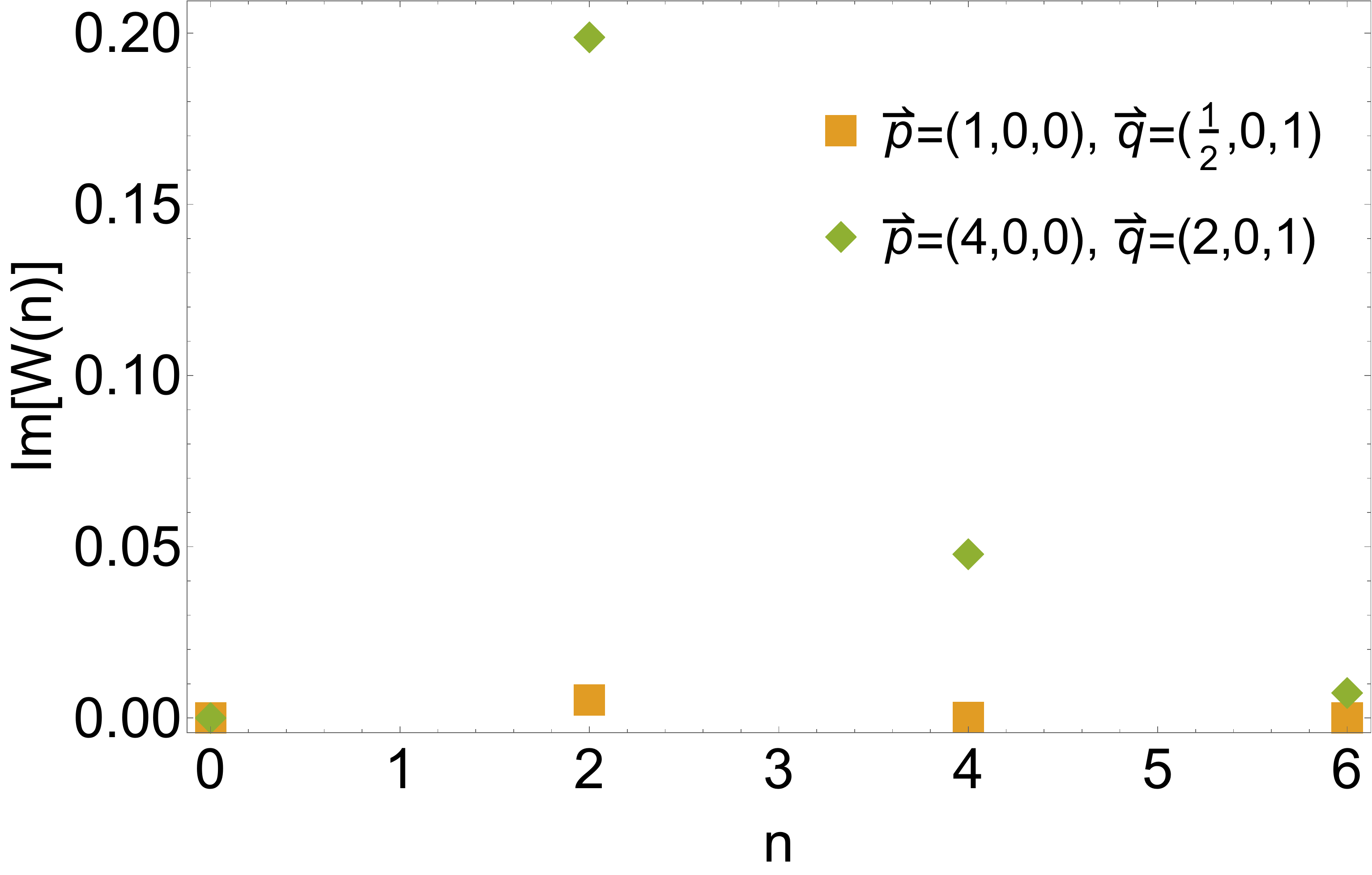}
\caption{We demonstrate the suppression factor obtained from the kinematic weight function $W(n)$ for several different kinematic choices. We examine the combinations $\mathbf{p}=(0,0,0)$, $\mathbf{q}=(0,0,1)$ (blue circles), $\mathbf{p}=(1,0,0)$, $\mathbf{q}=(1/2,0,1)$ (earth squares) and $\mathbf{p}=(4,0,0)$, $\mathbf{q}=(2,0,1)$ (garnet diamonds), where all momenta should be multiplied by 0.64~\si{GeV} to determine their physical values. As we explain, in the case where $\mathbf{p}\cdot\mathbf{q}\neq0$, the weighting function will be complex. Since $\mathbf{p}\cdot\mathbf{q}=0$ for the kinematic choice described by the blue circles, there is no imaginary part, and we thus exclude those points from the lower plot for clarity.}
\label{fig:weight}
\end{figure}

In this work, we are primarily interested in an extraction of the second Mellin moment of the pion's LCDA. From Eqn.~\ref{eq:hope}, it is possible to see that the $n$th moment is weighted by the kinematical factor 
\begin{equation}
\frac{\mathcal{C}_n^{2}(\eta)}{2^n(n+1)}C_W^{(n)}(\tilde{Q}^2)\zeta^n.
\end{equation}
For this section, we shall assume that the Wilson coefficients are unity. This results in $\mathcal{O}(\alpha_S)$ errors, but will not effect the features discussed here. We define the weight function
\begin{equation}
W(n)=\frac{\mathcal{C}_n^{2}(\eta)}{2^n(n+1)}\zeta^n.
\end{equation}
%
This weighting factor is the origin of the difficulty in extracting the higher moments in OPE approaches. For example in our numerical work we fix the physical size of the system to be $L \times a = 1.92~\si{fm}$ for all choices of the lattice spacing, $a$. Thus the smallest unit of momentum is
 \begin{equation}
 \Delta p=\frac{2\pi}{L a}=0.64~\si{GeV},
 \end{equation}
%
%
with the pion at rest $\mathbf{p}=(0,0,0)\times 0.64$~\si{GeV} and the current insertion momentum $\mathbf{q}=(0,0,1)\times 0.64$~\si{GeV} with $m_\pi=0.56$~\si{GeV} and $m_\Psi=2.7$~\si{GeV}, we find when scanning over $q_4$
\begin{align}
\max [W(0),q_4]&=1
\\
\max [W(2),q_4]&=0.008,
\end{align}
with higher moments further suppressed. Under normal circumstances, an extraction of even the first non-trivial (ie, the second) moment with this particular choice of kinematics would be a challenging task. Note however that by changing the kinematics, one may reduce the kinematic suppression. This fact is demonstrated in Fig.~\ref{fig:weight}, where a number of different choices of kinematics are shown. We note that in general the extraction of higher Mellin moments requires higher pion momentum, which poses a challenge for numerical determinations. 

Since we wish to numerically simulate the Compton tensor so that we may determine the second Mellin moment, it is advantageous to explore our kinematic options to best optimize the desired signal. In particular, by studying the properties of this weight function, we can determine kinematics which allow us direct access to the second Mellin moment, somewhat bypassing the kinematical suppression.

To begin, we note that while this function is real in Minkowski space, it is in general complex in Euclidean space. This is because $p_4=iE_\pi(\mathbf{p})$, and we take $q_4$ real. Noting again the definitions of the kinematical variables
\begin{align}
\eta&=\frac{p\cdot q}{\sqrt{p^2 q^2}},
\\
\zeta&=\frac{\sqrt{p^2 q^2}}{\tilde{Q}^2}.
\end{align}
We see that while $\zeta^{2n}$ is always real, under certain kinematical choices, $\eta$ is complex:
\begin{equation}
\eta=\frac{\mathbf{p}\cdot \mathbf{q}}{\sqrt{p^2 q^2}}+i\frac{E_\pi(\mathbf{p}) q_4}{\sqrt{p^2 q^2}}.
\end{equation}
Note that only the even moments are non-zero due to our assumption of isospin symmetry. The corresponding Gegenbauer polynomials are also even. Thus we only have even factors of $\eta$, and so we see that if we have the spatial inner product $\mathbf{p}\cdot \mathbf{q}\neq0$, the coefficient of $\expval{\xi^2}$ is complex. Note that since the kinematic factors are absent from the zeroth moment, this allows one separate the contribution from the lowest moment, and thus gain direct access to the second moment.

\begin{figure}
\centering
\includegraphics[width=\columnwidth]{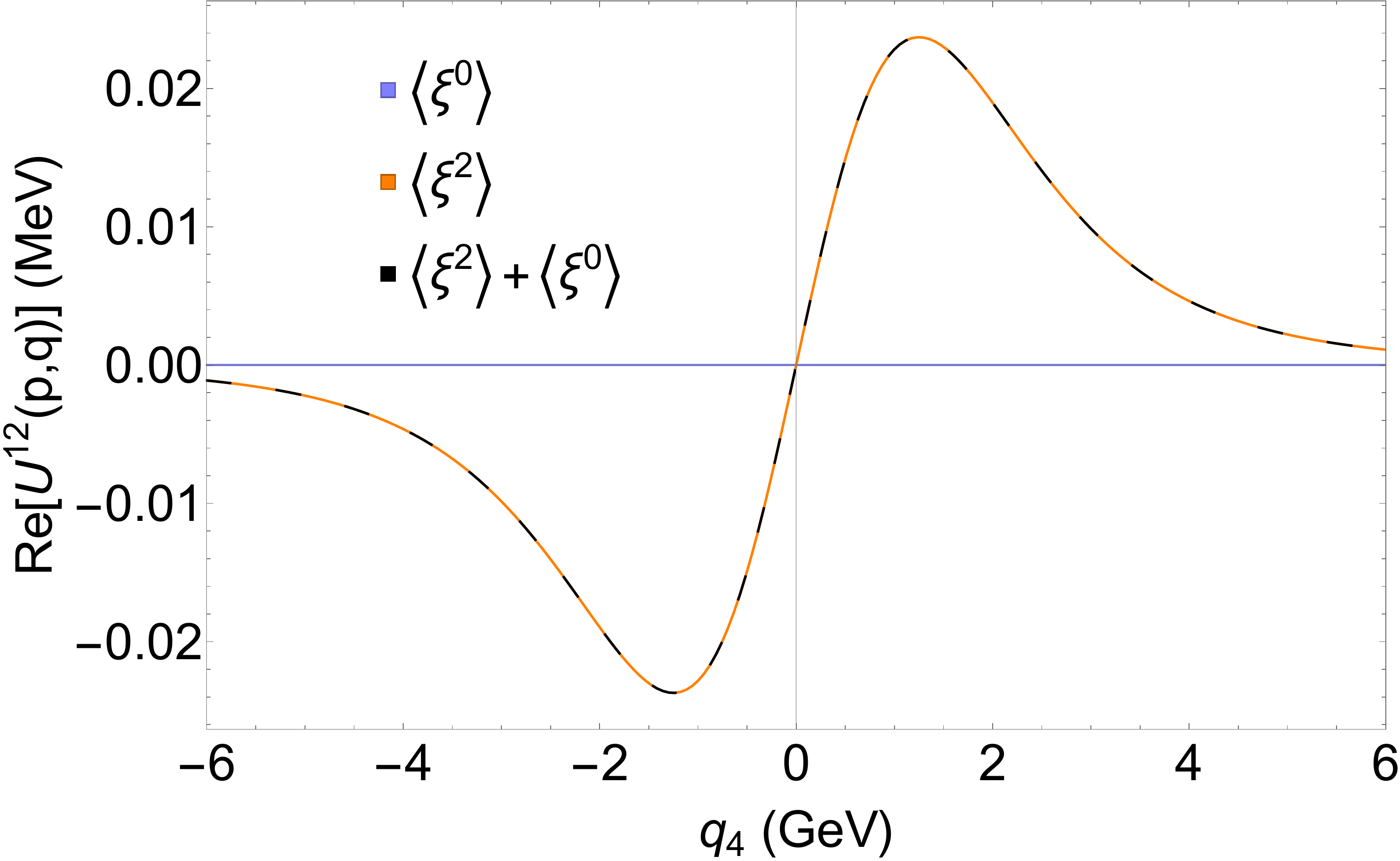}
\includegraphics[width=\columnwidth]{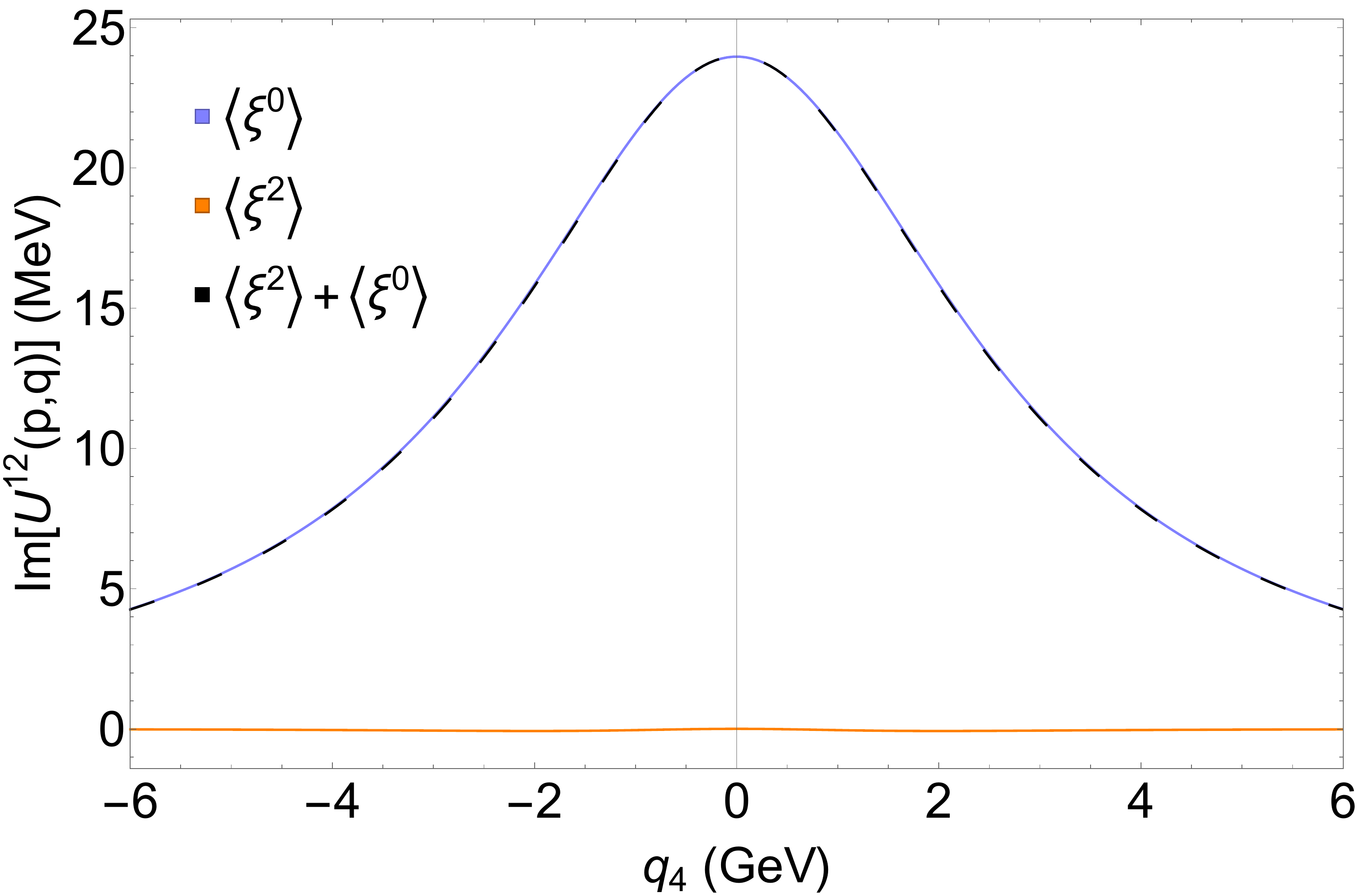}
\caption{Examining the special kinematics. By choosing the kinematics $\mathbf{p}=(1,0,0)\times 0.64$~\si{GeV}, $\mathbf{q}=(1/2,0,1)\times 0.64$~\si{GeV} and considering the real and imaginary parts independently, it is possible to see that while the imaginary part is saturated with the contribution from the lowest moment, the real part allows one to directly access the second Mellin moment directly.}
\label{fig:special-kinematics}
\end{figure}

There are several caveats to this. Firstly, the overall normalization of the HOPE can spoil this result. In particular consider the term $\epsilon^{\mu\nu\alpha\beta}q_\alpha p_\beta$. In this work, we study the combination $\mu=1$, $\nu=2$. We thus have
\begin{equation}
\epsilon^{12\alpha\beta}q_\alpha p_\beta=q_3 iE_\pi(\mathbf{p})-q_4p_3.
\end{equation}
Since this is an overall multiplicative factor, it will in general imbue all the moments (including the zeroth), with a complex kinematical factor. We can ensure this does not occur by taking kinematics where either $p_3=0$ or $q_3=0$. In either case, the kinematic factor will again be either purely real or purely imaginary, and thus the special kinematics may be used to directly access the second moment. Secondly, in this discussion we have neglected the role of the Wilson coefficients, however, we note that these can only give corrections which are $\mathcal{O}(\alpha_S)$, and will be numerically small. Thus as we shall see the `special kinematics' are still effective in isolating the second Mellin moment. A demonstration of the special kinematics is shown in Fig.~\ref{fig:special-kinematics}. To summarize, in this work, we use the conditions
\begin{align}
\mathbf{p}\cdot\mathbf{q}&\neq0,
\\
p_3&=0.
\end{align}
In particular, we choose to perform the simulations with the momentum 
\begin{align}
\mathbf{p}&=(1,0,0)\times 0.64~\si{GeV},
\\
\mathbf{q}&=(1/2,0,1)\times 0.64~\si{GeV},
\end{align}
The reason for the apparent fractional lattice momentum is that as we shall see the `physical' momenta are linear combinations of the inserted momenta, and in particular, we will see that we must include a factor of half in the definition of $\mathbf{q}$. This kinematic choice leads to less kinematical suppression;
\begin{align}
\max [W(0),q_4]&=1
\\
\max [W(2),q_4]&=0.02,
\end{align}
but most importantly allows one direct access to the second Mellin moment. Having now optimized our kinematical choice, we proceed to discuss the numerical simulation, and resulting extraction of the Mellin moment.

\section{Lattice Computation}
\begin{figure}[h]
  \centering
  \includegraphics[width=\columnwidth,trim={0 2cm 0 0},clip]{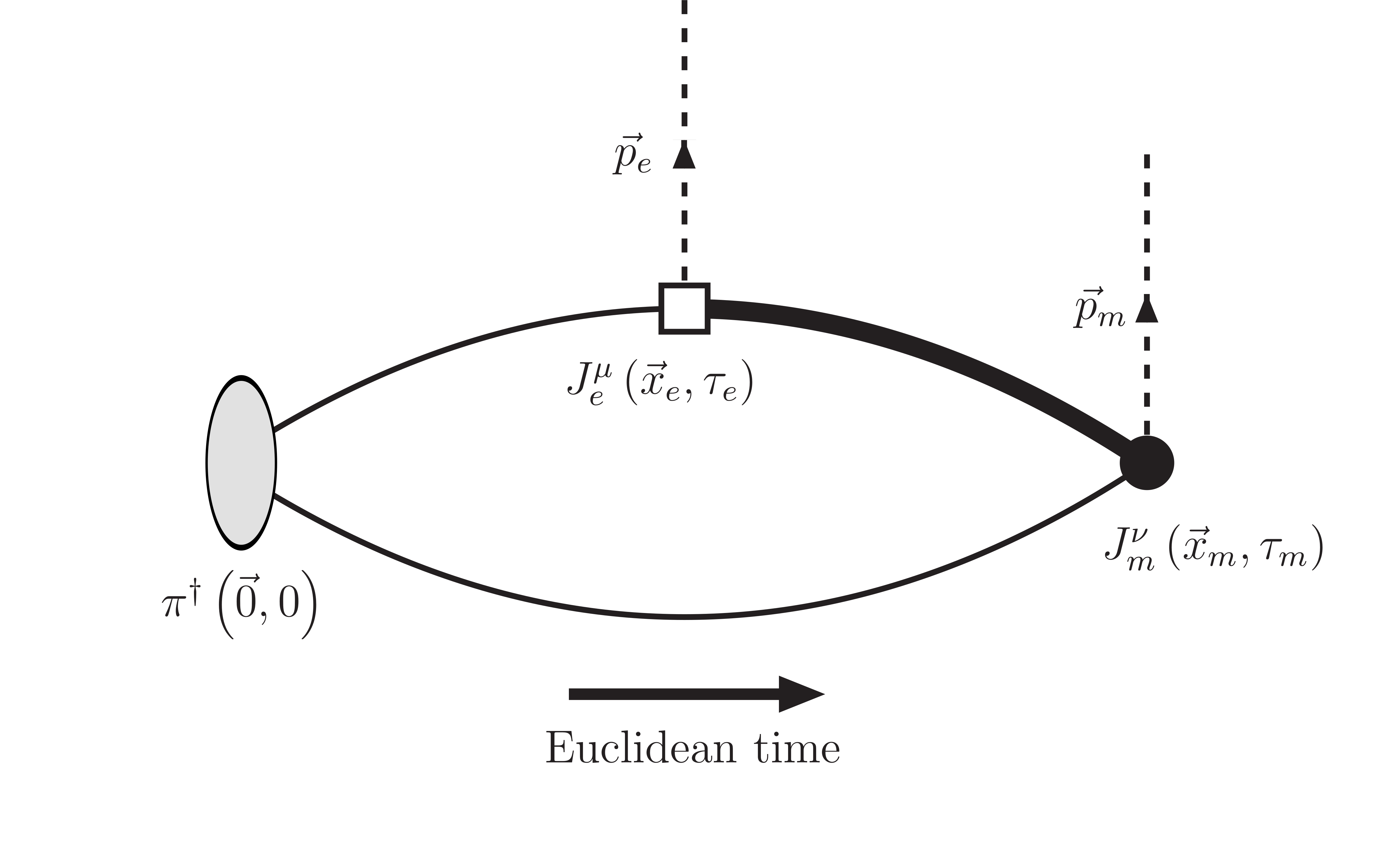}
  \caption{The three-point correlation function used to compute the hadronic tensor of the pion.  The pion is created at the origin and flavor-changing axial currents are inserted at times $\tau_e, \tau_m$.  The quark propagating between the currents is artifically heavy due to the flavor-changing nature of the currents.}
  \label{3pt-figure}
\end{figure}

The hadronic tensor is the Fourier transform of a current-current correlator, so it can be written in terms of 2- and 3-point functions of the form
\begin{equation}
  C_2 (\tau_\pi, \mathbf{p}) = \int d^3 \mathbf{x} \, e^{i\mathbf{p}\cdot \mathbf{x}} \langle 0 | \mathcal{O}_\pi (\mathbf{x}, \tau_\pi) \mathcal{O}_\pi^\dagger (\mathbf{0}, 0) | 0 \rangle
\label{2pt}
\end{equation}
\begin{equation}
  \begin{split}
    C_3^{\mu\nu} (\tau_e, \tau_m; \mathbf{p}_e, \mathbf{p}_m) = \int d^3 \mathbf{x}_e \, d^3 \mathbf{x}_m \, e^{i\mathbf{p}_e \cdot \mathbf{x}_e} e^{i\mathbf{p}_m \cdot \mathbf{x}_m} \\
    \langle 0 | \mathcal{T} \left[ J_A^\mu (\mathbf{x}_e, \tau_e) J_A^\nu (\mathbf{x}_m, \tau_m) \mathcal{O}_\pi^\dagger (\mathbf{0}, 0) \right] | 0 \rangle
  \end{split}
  \label{3pt}
\end{equation}
where $\mathcal{T}$ is the time-ordering operator, $\mathcal{O}_\pi$ is the pion interpolating operator, and
\begin{equation}
  J_A^\mu \equiv \bar \Psi \gamma^\mu \gamma^5 \psi + \bar \psi \gamma^\mu \gamma^5 \Psi
  \label{axial}
\end{equation}
is the flavor-changing axial current insertion operator that converts the pion's light quarks $\psi$ into valence heavy quarks $\Psi$ and vice versa.  In the large-time limit, the two- and three-point functions asymptote to
\begin{equation}
  C_2 (\tau_\pi, \mathbf{p}) \sim \frac{\left|\langle \pi(\mathbf{p})| \mathcal{O}^\dagger_\pi (\mathbf{0}, 0) | 0 \rangle\right|^2}{2E_\pi} e^{-E_\pi \tau_\pi}
  \label{2pt-fit}
\end{equation}

\begin{equation}
  \begin{split}
    C_3 (\tau_e, \tau_m; \mathbf{p}_e, \mathbf{p}_m) \sim \frac{\langle \pi(\mathbf{p}) | \mathcal{O}_\pi^\dagger (\mathbf{0}, 0) | 0 \rangle}{2E_\pi} e^{-E_\pi (\tau_e + \tau_m)/2} \\
    \int d^3 \mathbf{z} \, e^{i\mathbf{q} \cdot \mathbf{z}} \langle 0 | \mathcal{T}\left[ J_A^\mu \left( \frac{z}{2} \right) J_A^\nu \left( -\frac{z}{2} \right) \right]| \pi(\mathbf{p}) \rangle
  \end{split}
  \label{3pt-fit}
\end{equation}
with 
\begin{align}
z&= x_e - x_m,\label{eq:kin1}
\\
\mathbf{p}&= \mathbf{p}_e + \mathbf{p}_m,\label{eq:kin2}
\\
\mathbf{q}&=\frac{1}{2}(\mathbf{p}_m - \mathbf{p}_e).\label{eq:kin3}
\end{align}
The 3-point correlator is shown graphically in Fig.~\ref{3pt-figure} and can be computed via a sequential propagator through the operator.  The source and sink of the 2-point function and the source of the 3-point function are constructed using both Gaussian and link smearing to suppress excited state contamination.  Fitting $C_2$, $C_3^{\mu\nu}$ at large $\tau_\pi$, $\tau_e$, and $\tau_m$ lets us extract
\begin{equation}
  R^{\mu\nu} (\tau; \mathbf{p}, \mathbf{q})\!=\!\int d^3 \mathbf{z} \, e^{i \mathbf{q} \cdot \mathbf{z}} \langle 0 | \mathcal{T}\left[ J^\mu\left( \frac{z}{2} \right)J^\nu \left( -\frac{z}{2} \right) \right] | \pi(\mathbf{p}) \rangle
  \label{R-ratio}
\end{equation}
The hadronic tensor is then defined as the Fourier transform of $R^{\mu\nu}$ in the $\tau = z_4$ direction:
\begin{equation}
  U^{\mu\nu} (p,q) \equiv \int d\tau \, e^{iq_4 \tau} R^{[\mu\nu]}(\tau; \mathbf{p}, \mathbf{q})
  \label{hadronic-tensor-R}
\end{equation}

\subsection{Lattice Parameters}
In this study, we used Chroma \cite{chroma} to measure correlators at two heavy quark masses, with bare quark masses of about 1.6~\si{GeV} and 2.5~\si{GeV}.  (The renormalized quark masses found by the fits are substantially heavier than the bare masses.)  To accomodate such large quark masses, we use fine lattice spacings, ranging from 0.041~\si{fm} to 0.060~\si{fm} (and with future plans to include an additional ensemble with $a = 0.030$~\si{fm}).  The physical volumes are tuned to about 1.9~\si{fm} on all ensembles.

Due to critical slowing down, using dynamical fermions would be prohibitively expensive, especially for a preliminary simulation.  Therefore, this calculation is performed in the quenched approximation using lattices from Ref.~\cite{detmold-endres}.  The ensembles used here and the measurements performed on them are summarized in Table~\ref{ensembles}.


\begin{table*}
\caption{The ensembles used in this study and the number of measurements performed on each.}
\begin{ruledtabular}
\begin{tabular}{c c c c c c c c c c} 
$\beta$ & $a$ (fm) & $L^3 \times T$ &  $\kappa_\text{light}$ & $\kappa_\text{heavy, 1}$ & $\kappa_\text{heavy, 2}$ & $N_\text{cfg}$ & $N_\text{src}$ & Light Props & Heavy Props \\ \hline
6.30168 & 0.060 & $32^3 \times 64$ & 0.135146 & 0.119867 & 0.112779 & 450 & 7 & 3150 & 126,000 \\ 
6.43306 & 0.048 & $40^3 \times 80$ & 0.135170 & 0.122604 & 0.116599 & 250 & 2 & 500 & 20,000 \\
6.59773 & 0.041 & $48^3 \times 96$ & 0.135028 & 0.124420 & 0.119228 & 341 & 3 & 1023 & 40,920 \\ 
\end{tabular}
\end{ruledtabular}
\label{ensembles}
\end{table*}

We use Wilson clover fermions with the clover coefficient set non-perturbatively to the value in Ref.~\cite{csw}.  We tuned the pion mass to about 560~\si{MeV} across the ensembles.  In addition to reducing the computational cost, this unphysically heavy pion mass ensures that $m_\pi L > 5$ across our ensembles, suppressing finite-volume effects.

Note that in our calculational method, we need 40 heavy quark propagators per light quark propagator (2 heavy quark masses $\times$ 10 momentum insertions $\times$ 2 gamma matrices at current insertion).  However, each heavy quark inversion is substantially cheaper than each light quark inversion, so the large number of heavy quark inversions needed does not make the calculation intractable.

\subsection{Reducing Noise}
At the kinematics used, to $\mathcal{O}(\alpha_s)$, the second moment $\langle \xi^2 \rangle$ is proportional to the real part of the hadronic tensor (and the imaginary part of the hadronic tensor is mostly independent of $\langle \xi^2 \rangle$).  Thus, measuring $\text{Re}[U^{\mu\nu}]$ gives a clean probe of $\langle \xi^2 \rangle$ without much contamination from higher-twist effects.  However, while this is a clean signal, it is also a small one: At the kinematics used, the real part of $U^{\mu\nu}$ is 2--3 orders of magnitude smaller than the imaginary part.

The 3-point correlator (and therefore the ratio $R^{\mu\nu}$) is pure imaginary\footnote{For $p, q$ in Minkowski space, the hadronic tensor is purely imaginary, as can be seen from the operator product expansion.  The Euclidean-space data $R^{\mu\nu}$ are related to the Minkowski-space hadronic tensor via Laplace transform
$$ U^{\mu\nu}(q,p) = \int_{-\infty}^\infty d\tau\, e^{-q_0 \tau} R^{\mu \nu}(\tau; \mathbf{q}, \mathbf{p}) $$
whose kernel is purely real for real $q_0$.  Thus, if $U^{\mu\nu}$ with $q_0\in \mathbb{R}$ is imaginary, $R^{\mu\nu}(\tau; \mathbf{p},\mathbf{q})$ must be too.}, 
correspond to the antisymmetric and symmetric parts of $R^{\mu\nu}$.  Specifically,
\begin{align} 
  \text{Re}[U^{\mu\nu}&(\mathbf{p}, q)] = \text{Re} \left[ \int_{-\infty}^\infty d\tau \, R^{\mu\nu}(\tau; \mathbf{p}, \mathbf{q}) e^{-i q_4 \tau} \right] \nonumber \\ 
  &\propto \int_{0}^\infty d\tau \, \left[ R^{\mu\nu}(\tau; \mathbf{p}, \mathbf{q}) - R^{\mu\nu}(-\tau; \mathbf{p}, \mathbf{q}) \right] \sin (q_4 \tau)
  \label{real-part}
\end{align}
Thus, we need to measure a difference $R^{\mu\nu}(\tau; \mathbf{p}, \mathbf{q}) - R^{\mu\nu}(-\tau; \mathbf{p}, \mathbf{q})$ that is two orders of magnitude smaller than each of the terms constituting the difference.  The precision to which we can measure this difference depends on how well the two terms are correlated (which would cause correlated errors to cancel).  However, for moderately large $\tau$, the correlators $C_3^{\mu\nu}(\tau_e, \tau_e \pm \tau; \mathbf{p}_e, \mathbf{p}_m)$ used to compute $R^{\mu\nu}(\pm \tau; \mathbf{p}, \mathbf{q})$ have poorly correlated uncertainties since the sinks are temporally separated on the lattice.

\begin{figure}[h]
  \centering
  \includegraphics[width=\columnwidth]{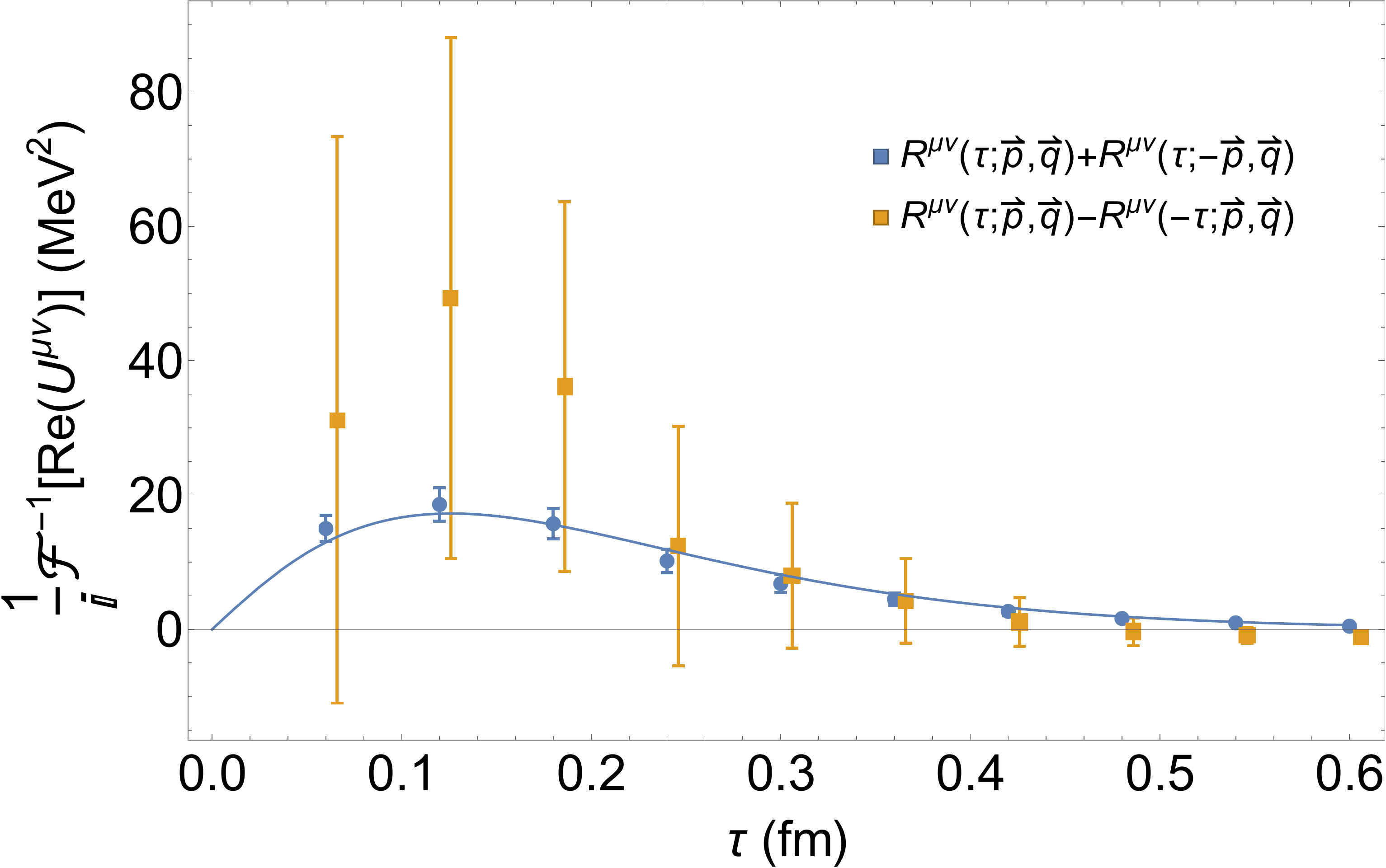}
  \caption{Comparing both sides of the equality in~\eqref{real-part-rewritten} $R^{\mu\nu}(\tau; \mathbf{p}, \mathbf{q}) + R^{\mu\nu}(\tau; -\mathbf{p}, \mathbf{q})$ (blue) and $R^{\mu\nu}(\tau; \mathbf{p}, \mathbf{q}) - R^{\mu\nu}(- \tau; \mathbf{p}, \mathbf{q})$ (earth), both  measured with two sources on 450 configurations.  These quantities agree in expectation, but the former has uncertainties an order of magnitude smaller than the latter.}
  \label{error-cancellation}
\end{figure}

We could obtain better correlations -- and therefore better error cancellation -- if we could compute $R^{\mu\nu}(-\tau; \mathbf{p}, \mathbf{q})$ using $C_3^{\mu\nu} (\tau_m, \tau_e; \mathbf{p}_e, \mathbf{p}_m)$, since then the correlators used to compute $R^{\mu\nu}(\tau; \mathbf{p}, \mathbf{q})$ and $R^{\mu\nu}(-\tau; \mathbf{p}, \mathbf{q})$ would be at the same timeslices (up to interchange of the two current insertions).  However, with the current setup where the sequential propagator passes through the first current inserted, this would require current insertions at all desired $\tau_m$, which would be prohibitively expensive.  Instead, we use $\gamma_5$-hermiticity to write
\begin{equation}
  C_3^{\mu\nu}(\tau_e, \tau_m; \mathbf{p}_e, \mathbf{p}_m)^* = C_3^{\nu\mu}(\tau_m, \tau_e; -\mathbf{p}_m, -\mathbf{p}_e)
  \label{gamma5-hermiticity}
\end{equation}
where $\mathbf{p}_e$ and $\mathbf{p}_m$ are related to $\mathbf{p}$ and $\mathbf{q}$ via \eqref{eq:kin2}, \eqref{eq:kin3}. This lets us compute both terms in the right-hand side of (\ref{real-part}) in terms of correlators with $\tau_m \geq \tau_e$, since
\begin{equation}
  R^{\mu\nu}(\tau; \mathbf{p}, \mathbf{q}) - R^{\mu\nu}(-\tau; \mathbf{p}, \mathbf{q}) = R^{\mu\nu}(\tau; \mathbf{p}, \mathbf{q}) + R^{\mu\nu}(\tau; -\mathbf{p}, \mathbf{q})
  \label{real-part-rewritten}
\end{equation}
Now, the terms in the right-hand side of (\ref{real-part-rewritten}) are more highly correlated, so we would expect larger cancellation of correlated errors.  This effect is shown in Fig.~\ref{error-cancellation}, where uncertainties are reduced by a factor of about 10 by using the right-hand side of (\ref{real-part-rewritten}) rather than the left-hand side.

\section{Fitting to the HOPE}
At the kinematics used here, the $n$th moment picks up a factor of $\left( \frac{p\cdot q}{\tilde{Q}} \right)^n \lesssim 0.12^n$, so the contribution of fourth moment is suppressed by a factor of about 50 relative to that of the second moment.  As a result, in this work, we will neglect higher-moment contributions, so we can write the operator product expansion as
\begin{equation}
  \begin{split}
    U^{\mu\nu} = \frac{2if_\pi \varepsilon^{\mu\nu\rho\lambda}q_\rho p_\lambda}{\tilde{Q}^2}\left[ \mathcal{C}_W^{(0)} + \langle \xi^2 \rangle \frac{6(p\cdot q)^2 - p^2 q^2}{6(\tilde{Q}^2)^2} \mathcal{C}_W^{\left( 2 \right)} \right. \\
    \left. + \cdots +\mathcal{O}\left( \frac{\Lambda_\text{QCD}}{\tilde{Q}} \right) \right]
  \end{split}
  \label{OPE}
\end{equation}
where $\tilde{Q}^2 = -m_\Psi^2 - q^2$, $m_\Psi$ is the renormalized heavy quark mass, and $\mathcal{C}_W^{(n)}$ are perturbatively calculable Wilson coefficients.  For this analysis, we have calculated the Wilson coefficients to 1-loop order, and we will publish the results in forthcoming work~\cite{inPreparationTheory}.  The remaining parameters ($f_\pi, m_\Psi, \langle \xi^2 \rangle$) will be fit to the data.  

In principle, one could measure $f_\pi$ separately using the pion-axial current.  However, measurements involving heavy quarks are known to involve additional normalization factors, which have been approximated by El-Khadra, Kronfeld, and Mackenzie~\cite{EKM}.  If we fit $f_\pi$ from the hadronic tensor, any errors in this overall normalization factor will be absorbed into $f_\pi$ (a nuisance parameter for us) rather than the parameter of interest, $\langle \xi^2 \rangle$.

It should be noted that the operator product expansion is a statement about continuum physics.  We will approximate the left-hand side of (\ref{OPE}) from an observable defined on the lattice and equate this to the right-hand side defined on the continuum, so our equation --- and therefore all parameters extracted from it --- are only accurate to $\mathcal{O}(a)$.  We will have to extrapolate away this lattice spacing dependence at the end of the calculation.

\begin{figure}[h]
\subfloat{
\includegraphics[width=\columnwidth]{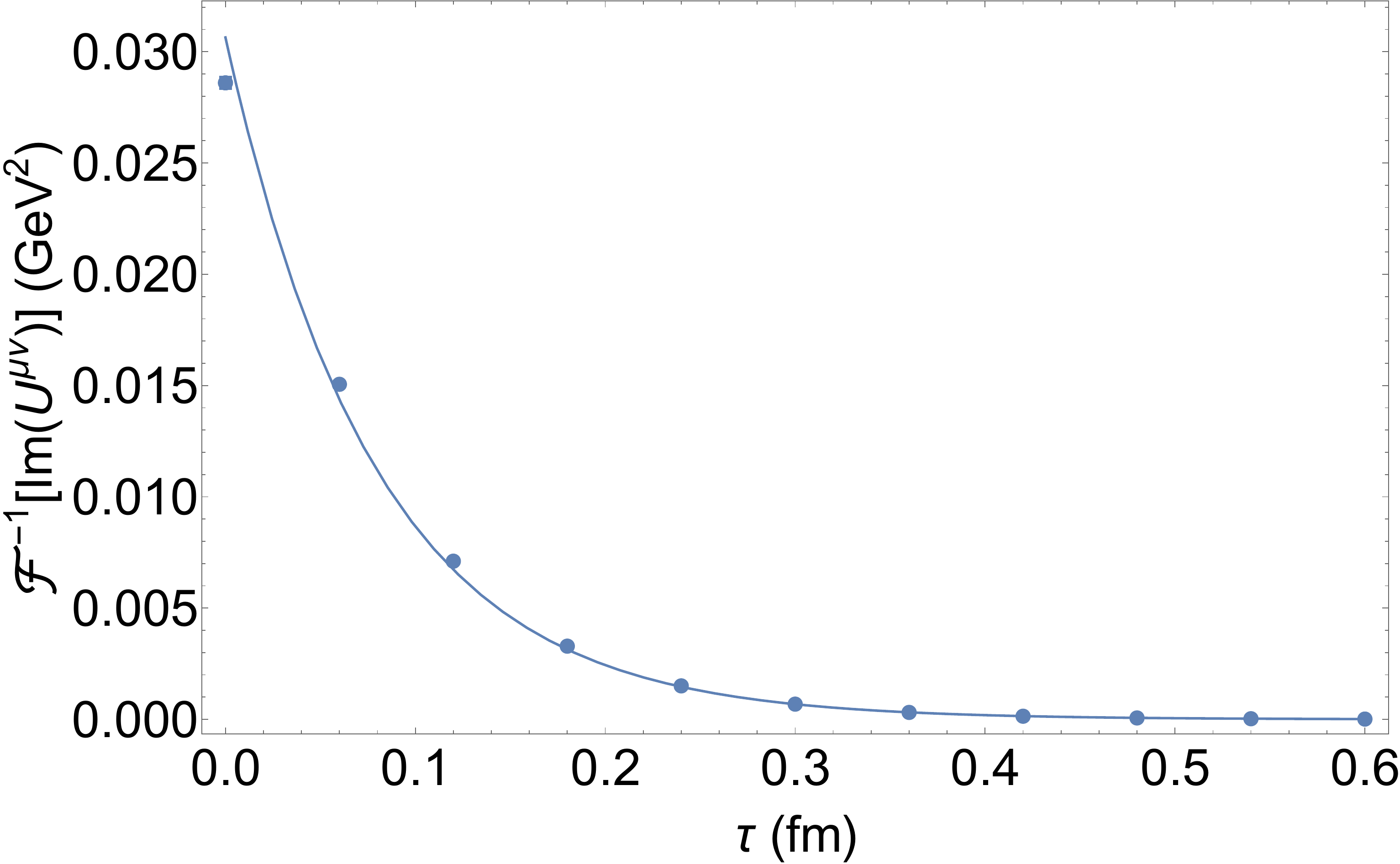}
}

\subfloat{
\includegraphics[width=\columnwidth]{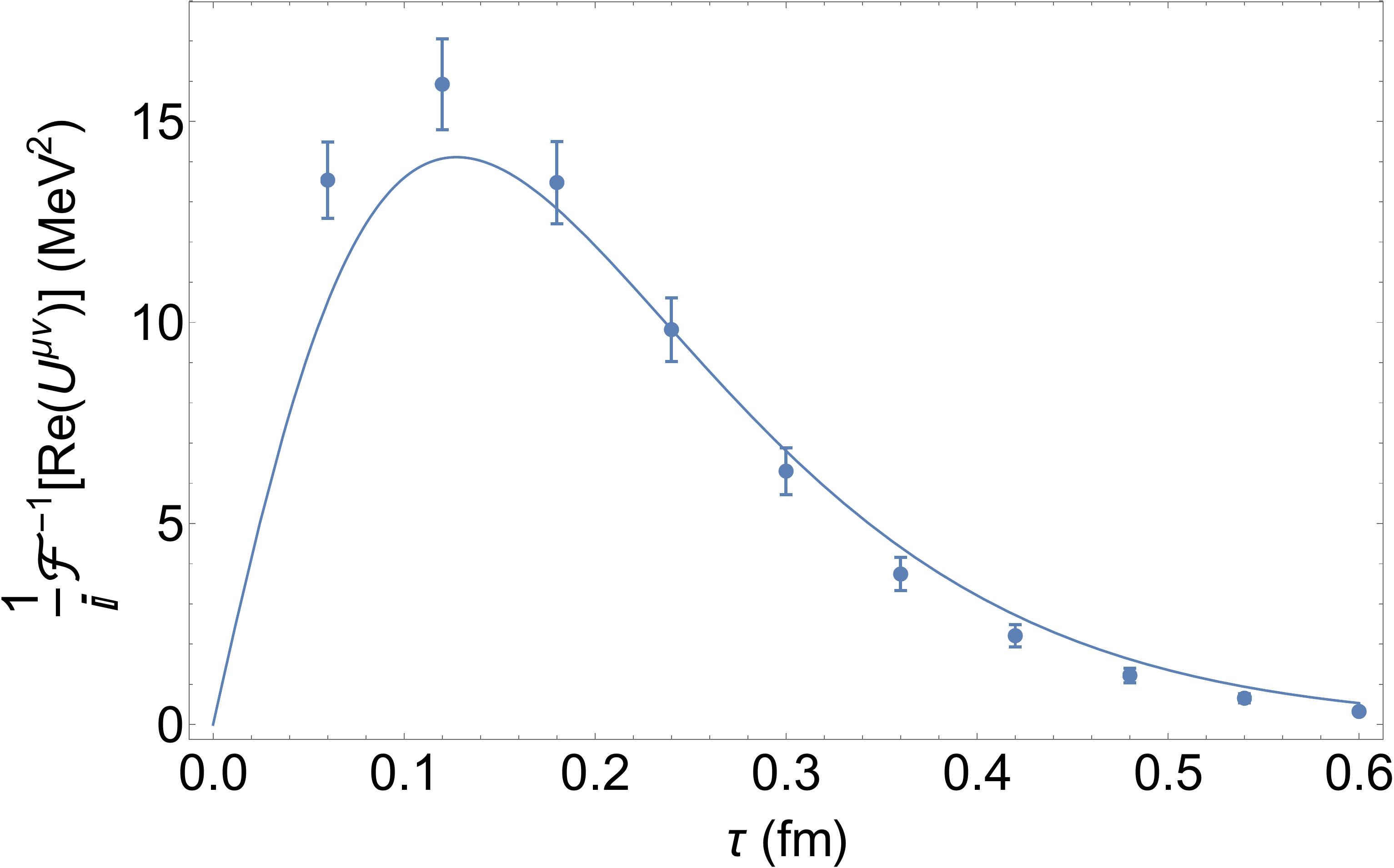}
}
\caption{The real and imaginary parts of the hadronic tensor data (at $a = 0.06$~\si{fm}) fitted to the continuum operator product expansion.}
\label{fit-results}
\end{figure}

To a good approximation, we can neglect the contribution of $\expval{ \xi^2}$ to the imaginary part of the hadronic tensor (this is a percent-level contribution) and use $\Im(U^{\mu\nu})$ to extract $f_\pi$ and $m_\Psi$. We can then use these fit values as inputs to the determination of $\expval{\xi^2}$ using $\Re(U^{\mu\nu})$. Fig.~\ref{fit-results} shows the results of this fitting procedure at $a = 0.06$~\si{fm}.


\section{Lattice Artifacts and Systematic Errors}
Our results (including those shown in Fig.~\ref{fit-results}) are contaminated with a variety of lattice artifacts and other systematic errors, including excited-state contamination and discretization effects.  We are also working at an unphysical pion mass of $m_\pi = 560$~\si{MeV} in the quenched approximation.  Finally, the operator product expansion we use is truncated, neglecting contributions both from higher moments (suppressed by the weight function $W(n)$) and from higher-twist effects that scale as $\Lambda_\text{QCD}/\tilde{Q}$ or $m_\pi/\tilde{Q}$.

Since this is a preliminary study, we will for now ignore the effects of quenching, unphysical $m_\pi$, and higher-twist effects. We also do not perform a chiral extrapolation, athough we note that others have studied the chiral behaviour of the Mellin moments, and found the chiral dependence to be small. For example, Refs.~\cite{Chen:2003fp,Chen:2005js} showed that at next-to-leading order in chiral perturbation theory, all possible non-analytic corrections to the matrix elements are contained in $f_\pi$. This has also been previously observed empirically, for example in Ref.~\cite{Braun:2006dg}. In this analysis, we will focus on excited-state contamination and lattice spacing dependence.

\subsection{Excited-State Contamination}
\begin{figure}
	\centering
	\includegraphics[width=\columnwidth]{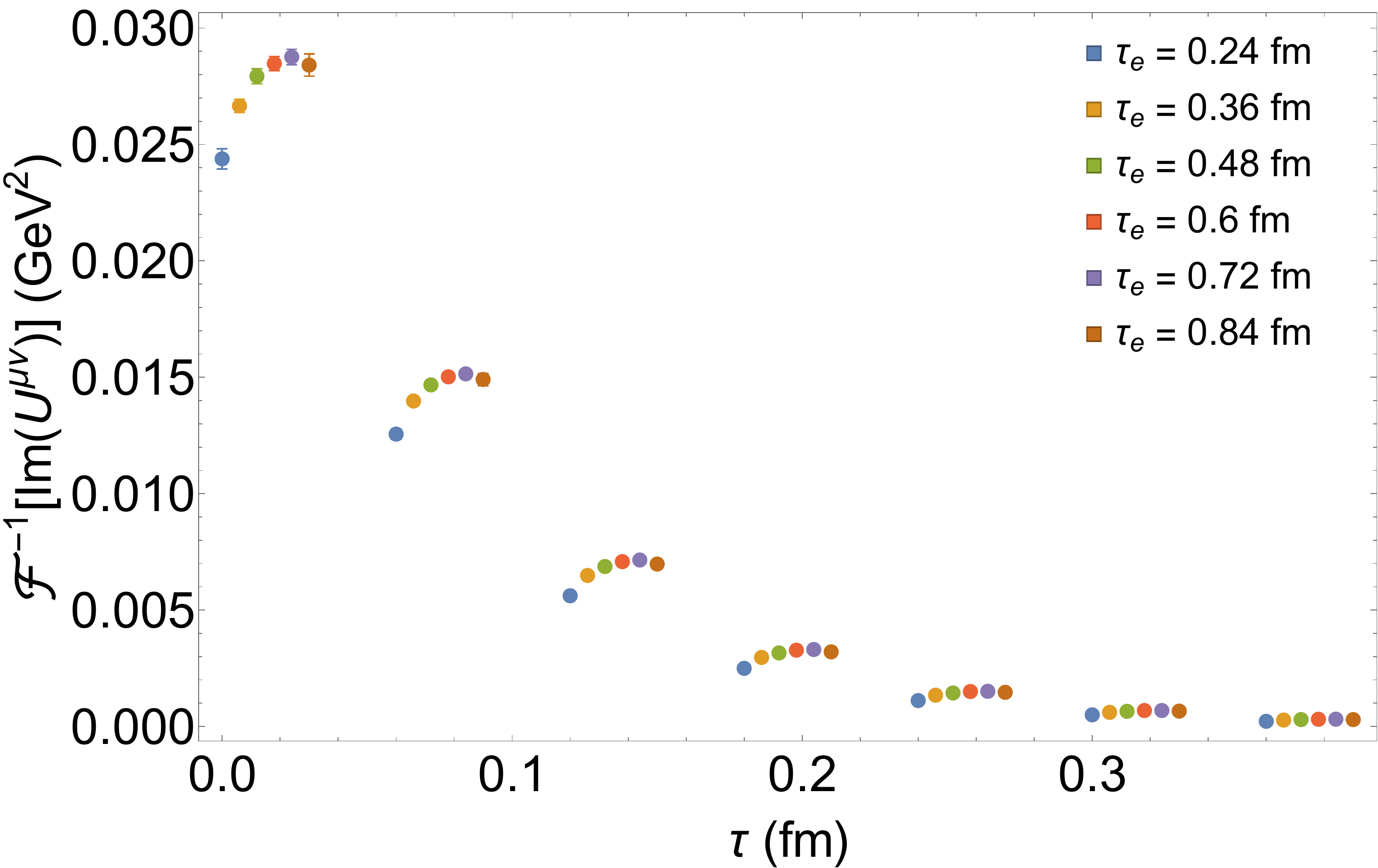}
  \caption{Excited state contamination is controlled by the separation between the pion source and the first current insertion ($\tau_e$).  The inverse Fourier transform of the hadronic tensor ($R^{\mu\nu}$) is measured at a range of $\tau_e$ values on the coarsest lattice ($a = 0.060$ fm).  Excited state contamination is clearly visible at $\tau_e \lesssim 0.4$ fm and is suppressed at larger $\tau_e$.}
	\label{excited-states-2}
\end{figure}

For equation (\ref{3pt-fit}) to be valid, the pion creation operator must be far from both current insertions.  For our measurements, $\tau_m \geq \tau_e$, so excited states should be suppressed exponentially in $\tau_e$.  Varying $\tau_e$ at a variety of current separations $\tau = \tau_m - \tau_e$ shows substantial contamination at small $\tau_e$ (see Fig.~\ref{excited-states-2}).

\begin{figure}
	\centering
	\includegraphics[width=\columnwidth]{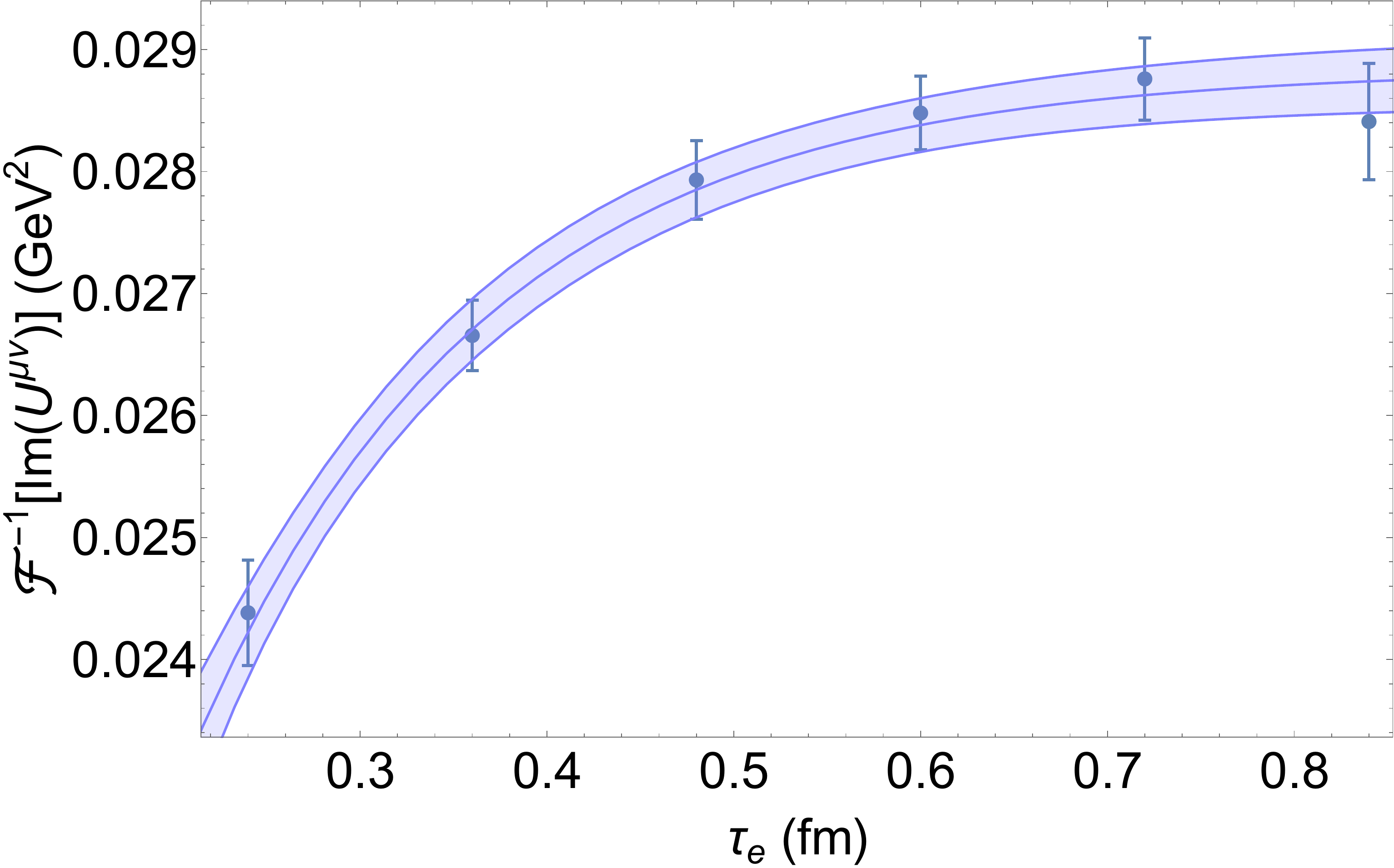}
  \caption{To make a quantitative estimate of excited-state contamination, we fit the $\tau=0$ data shown in Fig.~\ref{excited-states-2} with a two-state fit in $\tau_e$.  Excited-state contamination is about 1\% at $\tau_e = 0.7$ fm, which is smaller than many of the other statistical and systematic errors in this analysis.}
	\label{excited-states-3}
\end{figure}

In Fig.~\ref{excited-states-3}, we make a preliminary attempt to quantify this error.  A multi-state fit to the ratio $R^{\mu\nu}$ used to define the hadronic tensor shows that excited-state contamination effects are percent-level by $\tau_e \sim 0.7$ fm on the coarsest ensemble.  While it would be prohibitively expensive to rerun this measurement at such a wide range of $\tau_e$ on the finer ensembles, we estimate that excited-state contamination would be percent level there as well, which is smaller than the other errors (both statistical and systematic) in our analysis.

\subsection{Continuum Extrapolation}
Since our fit variables $f_\pi$ and $\langle \xi^2 \rangle$ were obtained by matching lattice data to the continuum OPE, they are contaminated with discretization errors.  While the fermions used have a non-perturbatively set clover coefficient, the axial current operators are unimproved, so the leading-order discretization artifacts are $\mathcal{O}(a)$.

\begin{figure}
\centering
\includegraphics[width=\columnwidth]{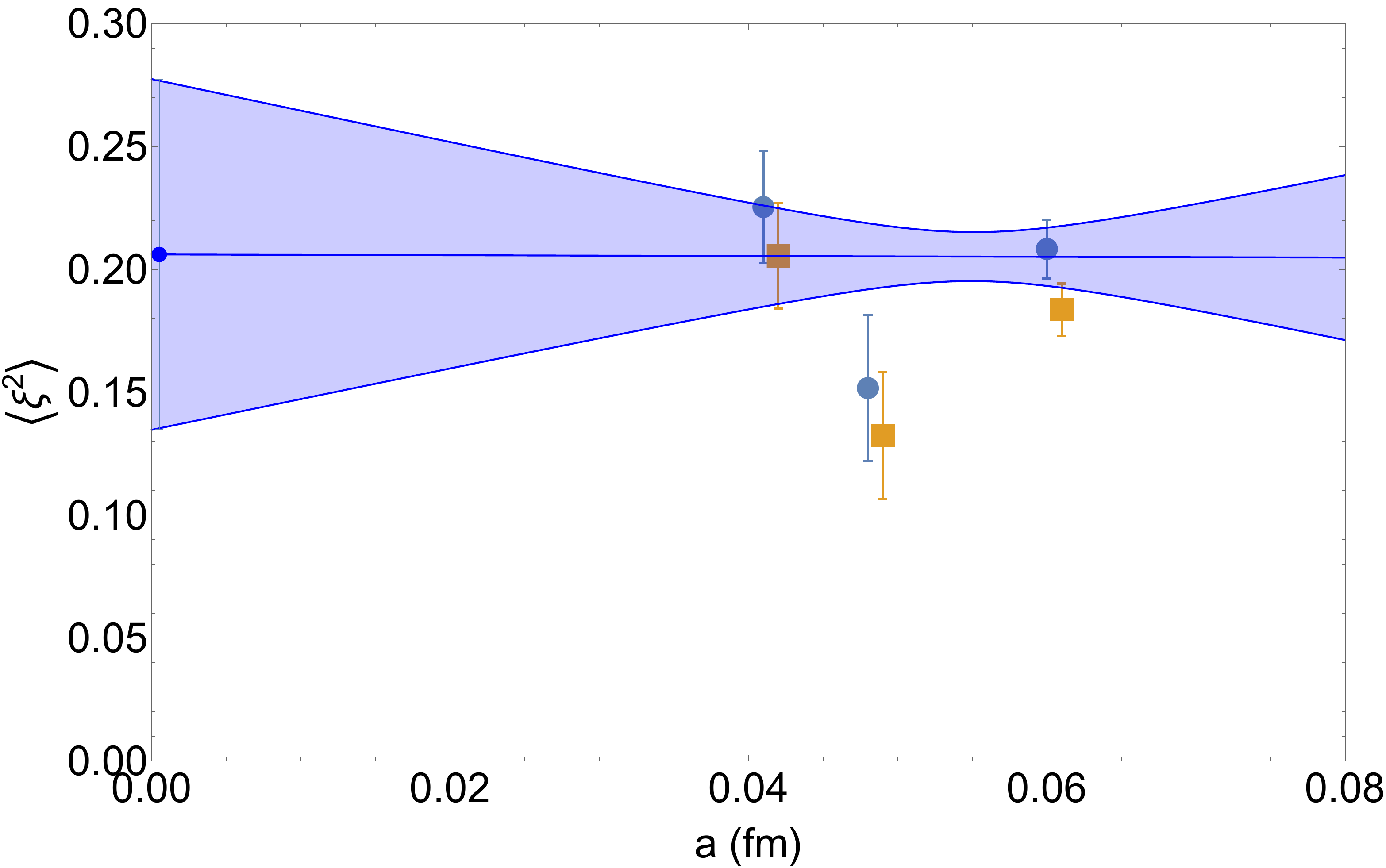}

\includegraphics[width=\columnwidth]{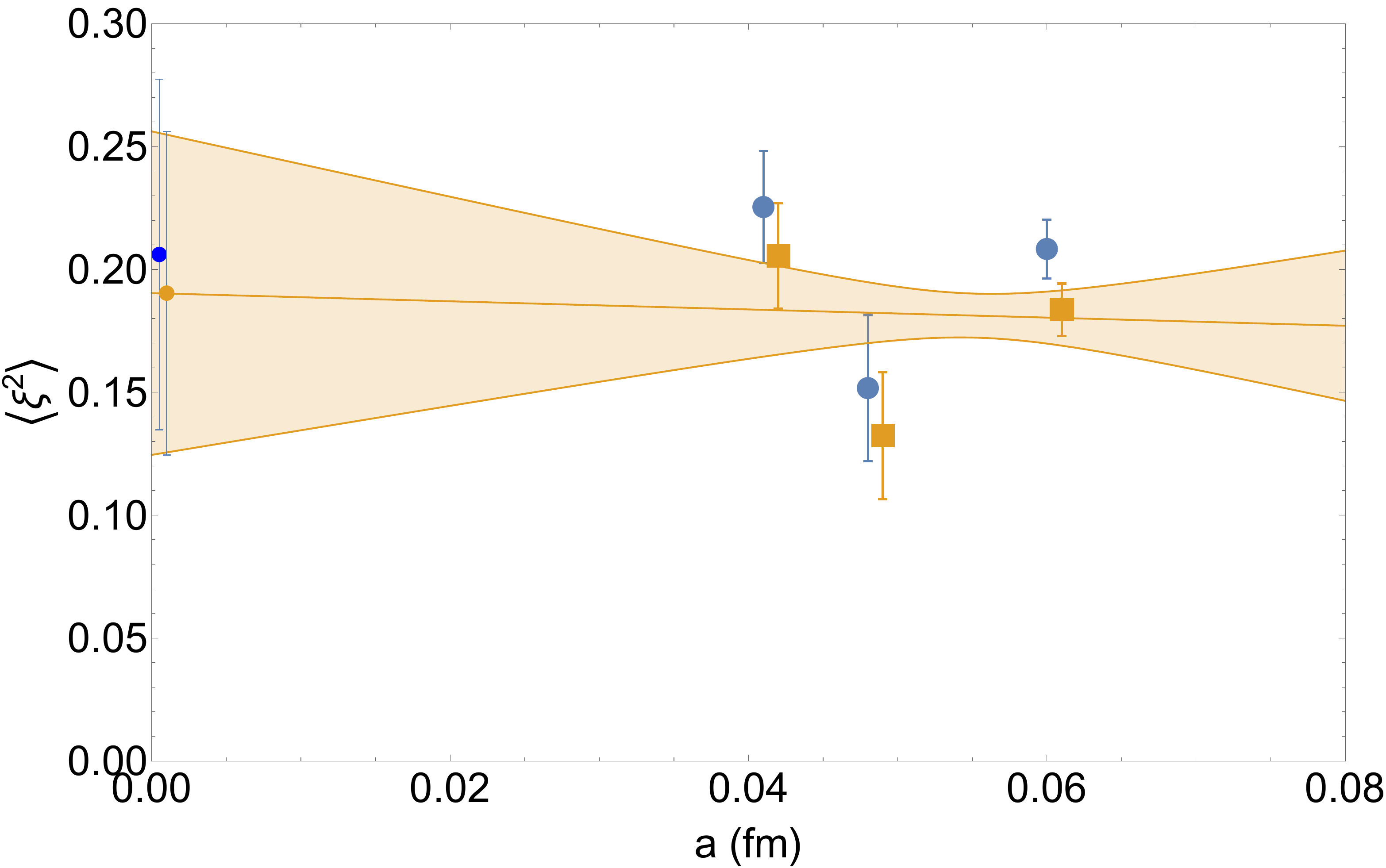}
  \caption{The continuum extrapolation of $\langle \xi^2 \rangle$ at heavy quark bare masses of about 1.6~\si{GeV} (blue) and 2.5~\si{GeV} (earth), corresponding to renormalized masses of about 2.6~\si{GeV} and 3.2~\si{GeV}, respectively.  The uncertainties in the continuum limit are large but should be reduced in future work.}
  \label{continuum-extrapolation}
\end{figure}

At each of the two heavy quark masses, the continuum limit is found by linear extrapolation of the form
\begin{equation}
    \langle \xi^2 \rangle_\text{latt} = \langle \xi^2 \rangle_\text{cont} + C a
\end{equation}
as shown in Fig.~\ref{continuum-extrapolation}.  Since this study is only preliminary, the range of lattice spacings used is currently small, leading to relatively large uncertainties in the continuum limit: the more precise of our measurements gives 
\begin{equation}
\langle \xi^2 \rangle_{\mu^2=2~\si{GeV}}^{\overline{\text{MS}}} = 0.19 \pm 0.07    
\end{equation}
Future work will increase statistics on the three existing ensembles and add a fourth, finer ensemble (at $a = 0.030$ fm) to approach the continuum limit.




The value of $\langle \xi^2 \rangle$ in the continuum should be independent of the heavy quark mass up to higher twist effects, which scale as inverse powers of $\tilde{Q}$ and therefore vanish as $m_\Psi \rightarrow \infty$.  The level of precision of current measurements is not sufficient to resolve these effects, so the fit values of $\langle \xi^2 \rangle$ are compatible at the two heavy quark masses in the continuum limit.

\section{Conclusion}
This preliminary study explores the potential of extracting the moments of the pion light-cone distribution amplitude using the operator product expansion with a heavy valence intermediate quark.  In the quenched approximation at an unphysical pion mass, this method does allow determination of $\langle \xi^2 \rangle$ in the continuum limit, albeit with a large uncertainty ($0.19 \pm 0.07$). Our ongoing work will significantly reduce the statistical and systematic uncertainties in this preliminary calculation.

Looking forward, this method can in principle be used to extract higher moments of the pion LCDA as well, provided we choose appropriate kinematics.  Future work will investigate the possibility of determining the fourth moment $\langle \xi^4 \rangle$ using this method.

\section*{Acknowledgements}

CJL and RJP were supported by the Taiwanese MoST Grant No. 109-2112-M-009-006-MY3 and MoST Grant No. 109-2811-M-009-516.  The work of IK is partially supported by the MEXT as ``Program for Promoting Researches on the Supercomputer Fugaku'' (Simulation for basic science: from fundamental laws of particles to creation of nuclei) and JICFuS. YZ is supported in part by the U.S. Department of Energy, Office of Science, Office of Nuclear Physics, under grant Contract Number DE-SC0012704 and within the framework of the TMD Topical Collaboration.  AVG is supported by the U.S. Department of Energy, Office of Science, Office of Nuclear Physics under grant Contract Numbers DE-SC0011090 and DE-SC0313021.  WD is supported by the U.S. Department of Energy, Office of Science, Office of Nuclear Physics under grant Contract Numbers DE-SC0011090 and DE-SC0018121.  Numerical calculations are performed at the HPC facilities at National Chiao-Tung University.  We acknowledge support of computing resources from ASRock Rack Inc. Finally, we all acknowledge the hospitality of National Chiao Tung University and Massachusetts Institute of Technology.




%
%
%
%

 \newpage
\bibliography{bibliography}

\end{document}